\documentclass[iop,apj,tighten]{emulateapj}

\usepackage{apjfonts}
\usepackage{graphicx}
\usepackage{dcolumn}
\usepackage{bm}
\usepackage{epsfig}

\usepackage{float}
\usepackage{amsmath,amstext}
\usepackage{floatflt}
\usepackage{subfigure}
\usepackage[usenames]{color}
\usepackage{ulem}
\usepackage{ragged2e}

\usepackage[breaklinks,colorlinks,citecolor=blue,linkcolor=magenta]{hyperref}
\usepackage[all]{hypcap} 

\shorttitle{Measuring the Hubble constant and spatial curvature}

\shortauthors{Park \& Ratra}

\begin{document}


\title{Measuring the Hubble constant and spatial curvature from supernova apparent magnitude, baryon acoustic oscillation, and Hubble parameter data}

\author{
Chan-Gyung Park\altaffilmark{1, 2} and
Bharat Ratra\altaffilmark{2}
}

\altaffiltext{1}{Division of Science Education and Institute of Fusion
                 Science, Chonbuk National University, Jeonju 54896, South Korea;
                 e-mail: park.chan.gyung@gmail.com}
\altaffiltext{2}{Department of Physics, Kansas State University, 116 Cardwell Hall,
                 Manhattan, KS 66506, USA
                 }

\date{\today}


\keywords{cosmological parameters --- large-scale structure of universe --- observations --- methods:statistical}

%
%
\begin{abstract}

Cosmic microwave background (CMB) anisotropy (spatial inhomogeneity) data 
provide the tightest constraints on the Hubble constant, matter
density, spatial curvature, and dark energy dynamics. Other data, sensitive
to the evolution of only the spatially homogeneous part of the cosmological 
model, such as Type Ia supernova apparent magnitude, baryon acoustic 
oscillation distance, and Hubble parameter measurements, can be used
in conjunction with the CMB data to more tightly constrain parameters.
Recent joint analyses of CMB and such non-CMB data indicate that slightly closed
spatial hypersurfaces are favored in nonflat untilted inflation models and 
that dark energy dynamics cannot be ruled out, and favor a smaller Hubble 
constant. We show that the constraints that follow from these non-CMB data 
alone are consistent with those that follow from the CMB data alone and so 
also consistent with, but weaker than, those that follow from the joint 
analyses of the CMB and non-CMB data.
\end{abstract}

\maketitle

%
%

\section{Introduction}

Establishing an accurate cosmological model that is consistent with 
observations is one of the primary goals of cosmology. Currently the 
spatially flat $\Lambda$CDM world model is considered to be the 
standard cosmological model \citep{Peebles1984}. In this model the current 
energy budget is dominated by the cosmological constant $\Lambda$ with
nonrelativistic cold dark matter (CDM) being the second biggest contributor. 
The former is responsible for the accelerated cosmological expansion at the 
present epoch, and the latter for initiating large-scale cosmological 
structure formation, and, along with nonrelativistic baryonic matter, for 
the earlier decelerated cosmological expansion. 

The spatially flat $\Lambda$CDM model is consistent with the cosmic microwave
background (CMB) anisotropy \citep{PlanckCollaboration2016, PlanckCollaboration2018}, Type Ia supernova 
apparent brightness \citep{Scolnicetal2018}, baryonic acoustic oscillation
distance \citep{Alametal2017}, and Hubble parameter
\citep{Farooqetal2017}\footnote{Hubble parameter measurements
provide evidence for the earlier nonrelativistic matter dominated 
cosmological expansion as well as the current dark energy powered 
accelerating cosmological expansion \citep{FarooqRatra2013, Farooqetal2013, Capozzielloetal2014, Morescoetal2016, Farooqetal2017, Yuetal2018, Jesusetal2018, Haridasuetal2018b}.} measurements. While CMB anisotropy data most tightly 
constrain cosmological models, at present no single kind of cosmological 
data is that restrictive, and it is the combination of CMB and non-CMB data that
results in powerful constraints (and breaks some of the degeneracy between 
correlated parameters).

Though the standard $\Lambda$CDM model assumes flat spatial hypersurfaces
and a constant dark energy density, current observations do not require 
either. To include nonzero spatial curvature in the analysis of CMB (and 
other spatial inhomogeneity) data requires the use of a nonflat inflation 
model to generate a physically consistent power spectrum of spatial 
inhomogeneities (as touched upon below and elsewhere; this has previously been 
ignored, resulting in invalid constraints on spatial curvature 
based on physically inconsistent power spectra). As an alternative to the 
constant dark energy density of $\Lambda$CDM model, the XCDM model is based
on a simple and widely used dynamical dark energy parameterization.
In this model, the ratio of dark energy pressure and density --- the equation of
state parameter ($w=p_X/\rho_X$) --- is constant. However, XCDM is not able 
to consistently describe the evolution of density inhomogeneities and so is 
not physically consistent. The $\phi$CDM model is a physically consistent 
dynamical dark energy model based on the evolution of a scalar field
\citep{PeeblesRatra1988,RatraPeebles1988}.\footnote{For earlier discussions
of cosmological constraints on the $\phi$CDM model see
\citet{Samushiaetal2007}, \citet{Yasharetal2009}, \citet{SamushiaRatra2010}, 
\citet{ChenRatra2011b}, \citet{Campanellietal2012}, 
\citet{Avsajanishvilietal2015}, \citet{Solaetal2017a}, \citet{Solaetal2017b}, 
\citet{Zhaietal2017}, \citet{Sangwanetal2018}, and references therein.}

Recently, \citet{Oobaetal2018a, Oobaetal2018b, Oobaetal2018c, Oobaetal2018d}
and \citet{ParkRatra2018a, ParkRatra2018b, ParkRatra2018c} reported
that compilations of Planck 2015 CMB anisotropy data 
\citep{PlanckCollaboration2016} and non-CMB data favor slightly closed
spatial hypersurfaces in the nonflat $\Lambda$CDM, XCDM, and 
$\phi$CDM dark energy untilted inflation
models, and noted that a dynamical dark energy density that varies both 
temporally and spatially cannot be ruled out.\footnote{This result 
differs from the \citet{PlanckCollaboration2016, PlanckCollaboration2018}
finding. As mentioned above, and discussed in detail elsewhere, the Planck 
analyses used a physically inconsistent power spectrum for energy density 
inhomogeneities, a physically inconsistent generalization of the 
nonflat untilted inflation model \citep{Gott1982, Hawking1984, Ratra1985} 
energy density inhomogeneity power spectrum 
\citep{RatraPeebles1995, Ratra2017}.} 

Most studies concentrate on using the most recent compilation of CMB
and non-CMB data to estimate the cosmological parameters as precisely as
possible. Here we want to examine the constraints on cosmological 
parameters that follow from the non-CMB observations alone, to avoid 
having to assume an energy density inhomogeneity power spectrum, and
to examine whether the non-CMB data constraints are consistent with 
the CMB ones.  
In this paper, we constrain the flat and nonflat $\Lambda$CDM,
XCDM, and $\phi$CDM dark energy models using an up-to-date collection of 
non-CMB data sets to constrain the spatially homogeneous cosmological models.
We use Type Ia supernova apparent magnitude, baryon acoustic 
oscillation distance, and Hubble parameter data to measure the 
matter density, Hubble constant, spatial curvature, and parameters
characterizing dark energy dynamics. We find that the conclusions obtained 
by jointly using CMB and non-CMB data sets, that favor slightly closed 
spatial hypersurfaces and a slightly smaller Hubble constant, and allow
for mild dark energy dynamics, also hold for the non-CMB data, but with lower 
statistical significance.\footnote{We emphasize that these results refer to
the cosmological parameter constraints, not to the goodness-of-fit of the 
best-fit set of cosmological parameters to the measurements. We find that the 
non-CMB data compilation we use here does not significantly distinguish between
any of the best-fit models on the basis of goodness-of-fit. When the CMB data 
are included in the mix we are unable to quantitatively determine the 
goodness-of-fit of the best-fit set of cosmological parameters to the 
measurements. This is in part due to the ambiguity in the 
number of degrees of freedom of the Planck CMB data 
\citep[see discussion in][]{Oobaetal2018a, Oobaetal2018b, Oobaetal2018c, Oobaetal2018d, ParkRatra2018a, ParkRatra2018b, ParkRatra2018c}. We also emphasize 
that qualitatively the slightly closed models better fit the lower multipole 
number CMB temperature anisotropy data and the weak lensing constraints on 
density inhomogeneities
\citep{DESCollaboration2018} while the flat models better fit the higher   
multipole number CMB temperature anisotropy data and the observed deuterium abundances 
\citep{Pentonetal2018}.}
  
In Sec.\ 2 the non-CMB data sets used in our analysis are briefly summarized.
In Sec.\ 3 we summarize our analysis methods that use the flat and nonflat 
$\Lambda$CDM, XCDM, and $\phi$CDM models. The observational constraints on 
the parameters of the six cosmological models are presented in Sec.\ 4. 
We summarize our results in Sec.\ 5.

\section{Data}

We use Type Ia supernova apparent magnitude (SN), baryon acoustic 
oscillation distance (BAO), and Hubble parameter [$H(z)$] measurements
to constrain the flat and nonflat $\Lambda$CDM, XCDM, and $\phi$CDM 
models.

We use the most recent SN data compilation, the Pantheon collection of
1048 Type Ia supernova apparent magnitude measurements over a redshift
range of $0.01 < z < 2.3$ \citep{Scolnicetal2018}.
This data set is a combination of Type Ia supernovae discovered by the 
Pan-STARRS1 Medium Deep Survey, the Sloan Digital Sky Survey, and the
Supernova Legacy Survey, together with
low-$z$ and Hubble Space Telescope SN samples.
In our analyses here we account for the statistical and systematic
uncertainties in the Pantheon measurements.   

We use a compilation of BAO data from 
\citet{Alametal2017}, \citet{Beutleretal2011}, \citet{Rossetal2015},
\citet{Ataetal2018}, \citet{Bautistaetal2017}, and \citet{Font-Riberaetal2014},
which is summarized in Table \ref{tab:bao}.
Here $D_M(z)$ is the comoving distance at redshift $z$, 
$D_H(z)=c/H(z)$, $D_V(z)=[cz D_M^2 (z) / H(z)]^{1/3}$, $D_A(z) = D_M(z)/(1+z)$, 
$r_d$ is the 
radius of the sound horizon at the drag epoch $z_d$, and $c$ is the speed of 
light (see Sec.\ 2.3 of \citealt{ParkRatra2018a}). Although 
Table \ref{tab:bao} here is similar to Table 1 of \citet{ParkRatra2018a},
here we exclude the growth rate ($f\sigma_8$) points from the 
Baryon Oscillation Spectroscopy Survey (BOSS) DR12 data \citep{Alametal2017}. Since we exclude these 
$f\sigma_8$ points, the DR12 covariance matrix between measurement errors
\citep{Alametal2017} we use here is
\begin{equation}
   \mathbf{C}_\textrm{DR12}=
   \begin{pmatrix}
      624.7  & 23.73  & 325.3 & 8.350 & 157.4 & 3.578  \\
      23.73  & 5.609  & 11.64 & 2.340 & 6.393 & 0.9681 \\
      325.3  & 11.64  & 905.8 & 29.34 & 515.3 & 14.10 \\
      8.350  & 2.340  & 29.34 & 5.423 & 16.14 & 2.853 \\
      157.4  & 6.393  & 515.3 & 16.14 & 1375  & 40.43 \\
      3.578  & 0.9681 & 14.10 & 2.853 & 40.43 & 6.259
   \end{pmatrix} .
\end{equation}
As in \citet{ParkRatra2018b, ParkRatra2018c} we also use the updated BAO 
data point of \cite{Ataetal2018}. In actual parameter estimation we use the 
probability distributions of the BAO data points of \citet{Rossetal2015} 
and \citet{Font-Riberaetal2014}, instead of the approximate
Gaussian constraints shown in Table \ref{tab:bao}.
See Sec.\ 2.3 of \citet{ParkRatra2018a} for more details about our 
procedure.

\begin{table}
\caption{BAO measurements.}
\begin{ruledtabular}
\begin{tabular}{ccc}
 $z_\textrm{eff}$                     &  Measurement                                          &   Reference    \\[+0mm]
 \hline \\[-2mm]
 $0.38$     & $D_M (r_{d,\textrm{fid}} / r_d)$ [Mpc]                   $= 1512.39 \pm24.99$   &  [1]  \\[+1mm]
 $0.38$     & $H (r_d / r_{d,\textrm{fid}})$ [km s$^{-1}$ Mpc$^{-1}$]  $= 81.21   \pm 2.37$   &  [1]  \\[+1mm]
 $0.51$     & $D_M (r_{d,\textrm{fid}} / r_d)$ [Mpc]                   $= 1975.22 \pm 30.10$  &  [1]  \\[+1mm]
 $0.51$     & $H (r_d / r_{d,\textrm{fid}})$ [km s$^{-1}$ Mpc$^{-1}$]  $= 90.90   \pm 2.33$   &  [1]  \\[+1mm]
 $0.61$     & $D_M (r_{d,\textrm{fid}} / r_d)$ [Mpc]                   $= 2306.68 \pm 37.08$  &  [1]  \\[+1mm]
 $0.61$     & $H (r_d / r_{d,\textrm{fid}})$ [km s$^{-1}$ Mpc$^{-1}$]  $= 98.96   \pm 2.50$   &  [1]  \\[+1mm]
 \hline \\[-2mm]
 $0.106$    & $r_d / D_V$                                              $= 0.327\pm0.015$  &  [2] \\[+1mm]
  \hline \\[-2mm]
 $0.15$     & $D_V (r_{d,\textrm{fid}} / r_d)$ [Mpc]                   $= 664\pm25$       & [3]  \\[+1mm]
  \hline \\[-2mm]
 $1.52$     & $D_V (r_{d,\textrm{fid}} / r_d)$ [Mpc]                   $= 3843\pm147$     & [4]  \\[+1mm]
   \hline \\[-2mm]
 $2.33$    &   $D_H^{0.7} D_M^{0.3} / r_d$                             $= 13.94\pm0.35$   &  [5] \\[+1mm]
   \hline \\[-2mm]
 $2.36$    & $D_H / r_d$                                               $= 9.0\pm0.3$      &  [6] \\[+1mm]
 $2.36$    & $D_A / r_d$                                               $= 10.8\pm0.4$     &  [6] \\[+0mm]
\end{tabular}
\\[+1mm]
References: [1] \citet{Alametal2017}, [2] \citet{Beutleretal2011}, [3] \citet{Rossetal2015}, [4] \citet{Ataetal2018}, [5] \citet{Bautistaetal2017}, [6] \citet{Font-Riberaetal2014}.
Note: The sound horizon size (at the drag epoch) of the fiducial model
is $r_{d,\textrm{fid}}=147.78~\textrm{Mpc}$ in \cite{Alametal2017} and \cite{Ataetal2018},
and $r_{d,\textrm{fid}}=148.69~\textrm{Mpc}$ in \cite{Rossetal2015}.
\end{ruledtabular}
\label{tab:bao}
\end{table}

For $H(z)$ data, we use the collection of 31 Hubble parameter measurements over a large 
redshift range ($0.070 \le z \le 1.965$) listed in Table 2 of 
\citet{ParkRatra2018a}. See \citet{Morescoetal2018} for a recent discussion of Hubble parameter measurement error bars.

\section{Methods}

We measure the parameters of the flat and nonflat $\Lambda$CDM, XCDM, and 
$\phi$CDM models by comparing model predictions with the observed SN apparent
 magnitudes, BAO distances, and Hubble parameters over a large range of 
redshift. 

The evolution of the spatially homogeneous background in the $\Lambda$CDM and 
XCDM models is usually described by the evolution of the Hubble parameter.
For the nonrelativistic matter and dark energy dominated epochs, the Hubble
parameter $H(a)$ as a function of the scale factor $a$ (normalized to
be unity now) is
\begin{equation}
   \left( \frac{H}{H_0} \right)^2 = \Omega_m a^{-3} + \Omega_k a^{-2} + \Omega_\Lambda
\label{eq:lcdm}
\end{equation}
for the $\Lambda$CDM model, and
\begin{equation}
   \left( \frac{H}{H_0} \right)^2 = \Omega_m a^{-3} + \Omega_k a^{-2} + \Omega_X a^{-3(1+w)}
\label{eq:xcdm}
\end{equation}
for the XCDM parameterization. Here $H_0$ is the Hubble constant, 
the nonrelativistic matter density parameter present value is the sum of 
present baryonic matter and CDM density parameters, $\Omega_m = \Omega_b + 
\Omega_c$, $\Omega_k$ is the present value of the spatial curvature density 
parameter, and $\Omega_\Lambda$ and $\Omega_X$ are the present values
of the dark energy density parameters in the $\Lambda$CDM and XCDM models, 
respectively. In the limit 
$w = -1$ the XCDM dark energy becomes the cosmological constant $\Lambda$. 

In the $\phi$CDM model we consider a minimally coupled dark energy scalar
field $\phi$ with an inverse power-law potential energy density
\begin{equation}
    V(\phi)=V_1 \phi^{-\alpha},
\end{equation}
where $\alpha$ is a positive constant parameter and $V_1$ is determined in 
terms of $\alpha$ \citep{PeeblesRatra1988}. In the limit $\alpha = 0$ the
scalar field dark energy becomes the cosmological constant $\Lambda$.

For the background evolution of the $\phi$CDM model in the nonrelativistic 
matter and scalar field dominated epochs we use
\begin{equation}
   \left(\frac{H}{H_0}\right)^2
     = \frac{1}{1-\frac{1}{6}\left(\phi' \right)^2} 
       \left[ \Omega_m a^{-3} + \Omega_k a^{-2}
       + \frac{1}{3}\hat{V}(\phi) \right] ,
\label{eq:qcdm}
\end{equation}
where the evolution of the dark energy scalar field is governed by the 
equation of motion
\begin{equation}
   \phi'' + \left( 3 + \frac{\dot{H}}{H^2} \right) \phi'
      + \hat{V}_{,\phi} \left(\frac{H_0}{H}\right)^2
    = 0 .
\label{eq:q-dyn}
\end{equation}
Here $\phi' \equiv d\phi / d\ln a$, $H=\dot{a} / a$,
$\hat{V}(\phi) \equiv V(\phi) / H_0^2$,
$\hat{V}_{,\phi}=-\hat{V}_1 \alpha \phi^{-\alpha -1}$,
$\hat{V}_1 \equiv V_1 / H_0^2$,
and an overdot denotes the time derivative $d/dt$.  We have chosen 
units such that the Newtonian gravitational constant $G \equiv 1/ 8\pi$.
We use the initial conditions of \citet{PeeblesRatra1988} at
scale factor $a_i=10^{-10}$. This places the homogeneous background scalar field
on the attractor/tracker solution \citep{PeeblesRatra1988, RatraPeebles1988,
Pavlovetal2013}.
For a given set of cosmological parameters and initial conditions for the
scalar field, we numerically determine the value of $\hat{V}_1$ 
to satisfy the condition $H/H_0=1$ at the present epoch 
(when $a = 1$). The current value of the dark energy density parameter is
$\Omega_\phi = (\phi_0')^2 / 6 + \hat{V}(\phi_0) / 3$, where $\phi_0$ and 
$\phi_0'$ are the current values of $\phi$ and $\phi'$.

The version of Eqs.\ (\ref{eq:lcdm})--(\ref{eq:q-dyn}) we use in the actual 
computations also take into account the contribution of photons and massless 
and massive neutrinos. We assume that the present CMB temperature 
$T_0=2.7255~\textrm{K}$, that the effective number of neutrino species
$N_\textrm{eff}=3.046$, and one massive neutrino species 
(with mass $m_\nu=0.06~\textrm{eV}$).

To obtain the likelihood distributions of the cosmological parameters, we use 
the Markov chain Monte Carlo (MCMC) method that randomly explores the parameter 
space based on the probability function $P(\mathbf{m} | \mathbf{d}) \propto \textrm{exp}(-\chi^2 /2)$,
where $\mathbf{m}$ and $\mathbf{d}$ denote model and data, respectively, and 
$\chi^2=\chi_\textrm{SN}^2 + \chi_\textrm{BAO}^2 + \chi_{H(z)}^2$ is the sum
of individual contributions from the SN, BAO, and $H(z)$ data.
When comparing the Pantheon SN apparent magnitude data with model 
predictions we use the $\chi_{\textrm{SN}}^2$ defined in Appendix C (Eq.\ C1) 
of \citet{Conleyetal2011}. The SN covariance matrix $\mathbf{C}_\textrm{SN}$ 
is the sum of the diagonal statistical uncertainty covariance matrix,
$\mathbf{D}_\textrm{stat}=\textrm{diag}(\sigma_{\textrm{SN},i}^2)$, and the 
systematic uncertainty covariance matrix,
$\mathbf{C}_\textrm{sys}$: $\mathbf{C}_\textrm{SN}=\mathbf{D}_\textrm{stat}+\mathbf{C}_\textrm{sys}$.
For the BAO data, $\chi_\textrm{BAO}^2$ is the sum of contributions
from each BAO measurement. For example for the BOSS DR12 BAO data we have
$\chi_\textrm{DR12}^2=\mathbf{X}^T \mathbf{C}_{\textrm{DR12}}^{-1} \mathbf{X}$ 
where $\mathbf{X}$ is a vector whose elements are the differences between model
predictions and data points (the first six entries in Table \ref{tab:bao}).
For Hubble parameter data, 
$\chi_{H(z)}^2=\sum_{i=1}^{31} [H(z_i)-H_\textrm{obs}(z_i)]^2/\sigma_{H(z),i}^2$.

We constrain the flat $\Lambda$CDM model with three cosmological parameters
($\Omega_b h^2$, $\Omega_c h^2$, $H_0$) and the nonflat $\Lambda$CDM model
with four parameters ($\Omega_b h^2$, $\Omega_c h^2$, $H_0$, $\Omega_k$),
where $h = H_0 / (100$ km s$^{-1}$ Mpc$^{-1}$). 
We add one more free parameter, the equation of state parameter $w$ for the 
XCDM parameterization, and the scalar field potential parameter $\alpha$ 
for the $\phi$CDM model.\footnote{
Although we use the parameter $\theta_{\textrm{MC}}$, the approximate angular
size of the sound horizon at recombination \citep{PlanckCollaboration2014}, 
instead of $H_0$ in our $\Lambda$CDM and XCDM model analyses, we 
instead record the derived $H_0$ as one of the main 
cosmological parameters for these models. For the $\phi$CDM model, 
however, $H_0$ (not $\theta_{\textrm{MC}}$) is the active parameter in the 
MCMC analysis.}

\begin{figure*}
\centering
\mbox{\includegraphics[width=87mm]{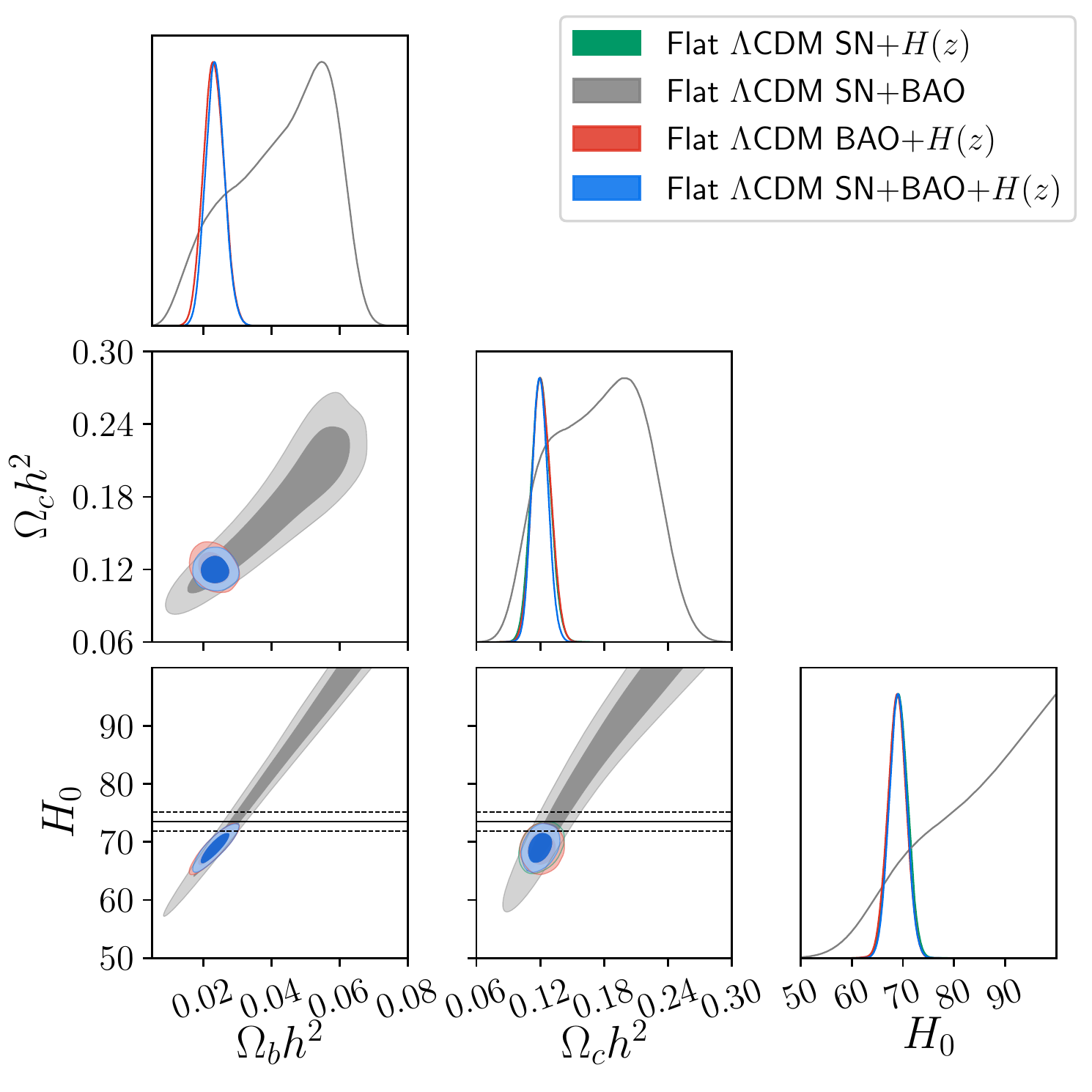}}
\mbox{\includegraphics[width=87mm]{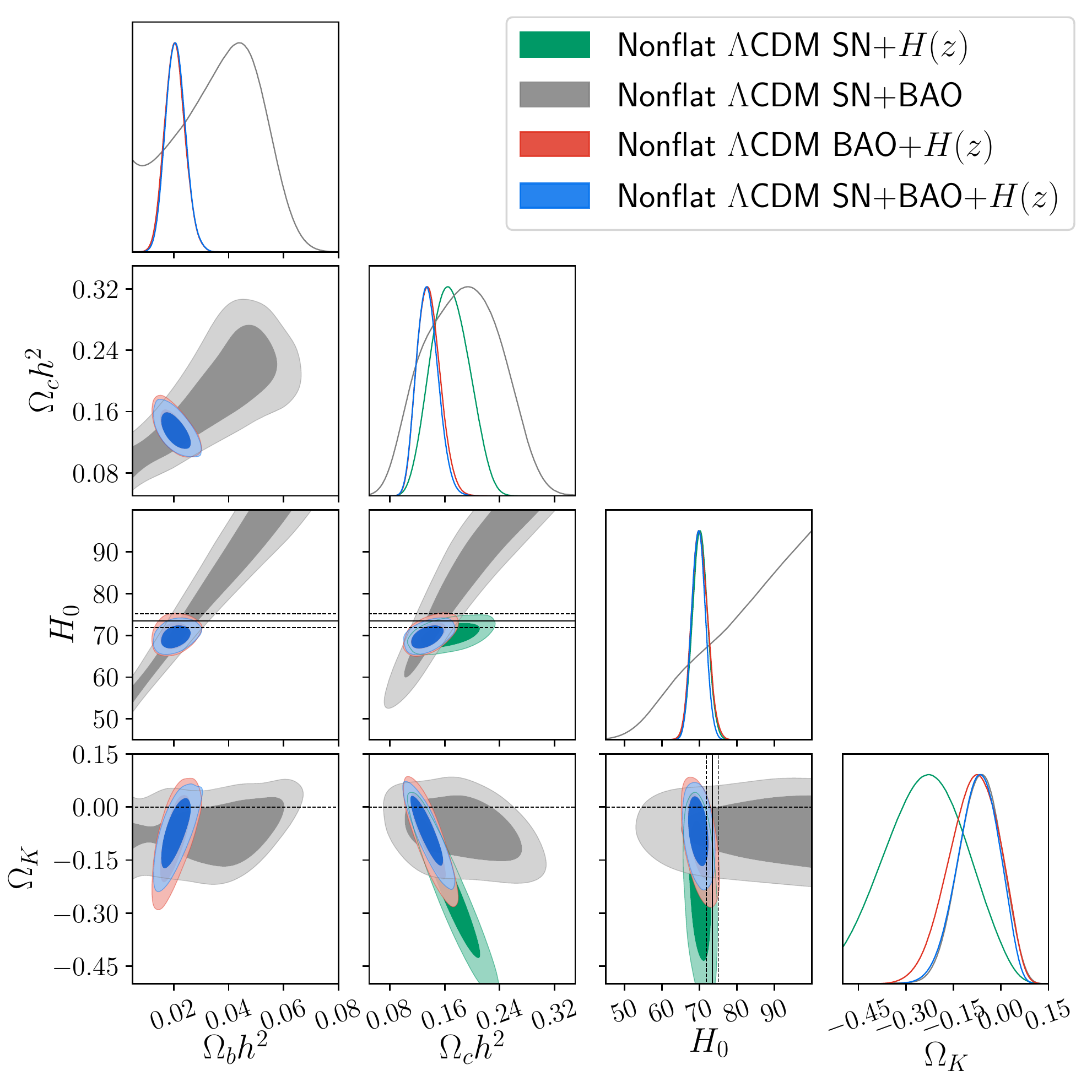}}
\caption{Left panel: two- and one-dimensional likelihood distributions of flat $\Lambda$CDM
model parameters ($\Omega_b h^2$, $\Omega_c h^2$, $H_0$) constrained using the SN+$H(z)$, SN+BAO, BAO+$H(z)$, and SN+BAO+$H(z)$ data combinations. 
Right panel: similar distributions of nonflat $\Lambda$CDM model parameters
($\Omega_b h^2$, $\Omega_c h^2$, $H_0$, $\Omega_k$).
Horizontal and vertical lines in the $H_0$-related plots indicate the recent
local Hubble constant measurement (solid lines) and 68.3\% confidence limits
(dashed lines) of \citet{Riessetal2018},
$H_0 = 73.48 \pm 1.66$ $\textrm{km} \textrm{s}^{-1} \textrm{Mpc}^{-1}$. 
The dashed lines in $\Omega_k$-related plots demarcate the spatially-flat model.
In both panels, the baryonic density parameter $\Omega_b h^2$ is not 
constrained by the SN+$H(z)$ data.
}
\label{fig:para_LCDM}
\end{figure*}

\begin{figure*}
\centering
\mbox{\includegraphics[width=87mm]{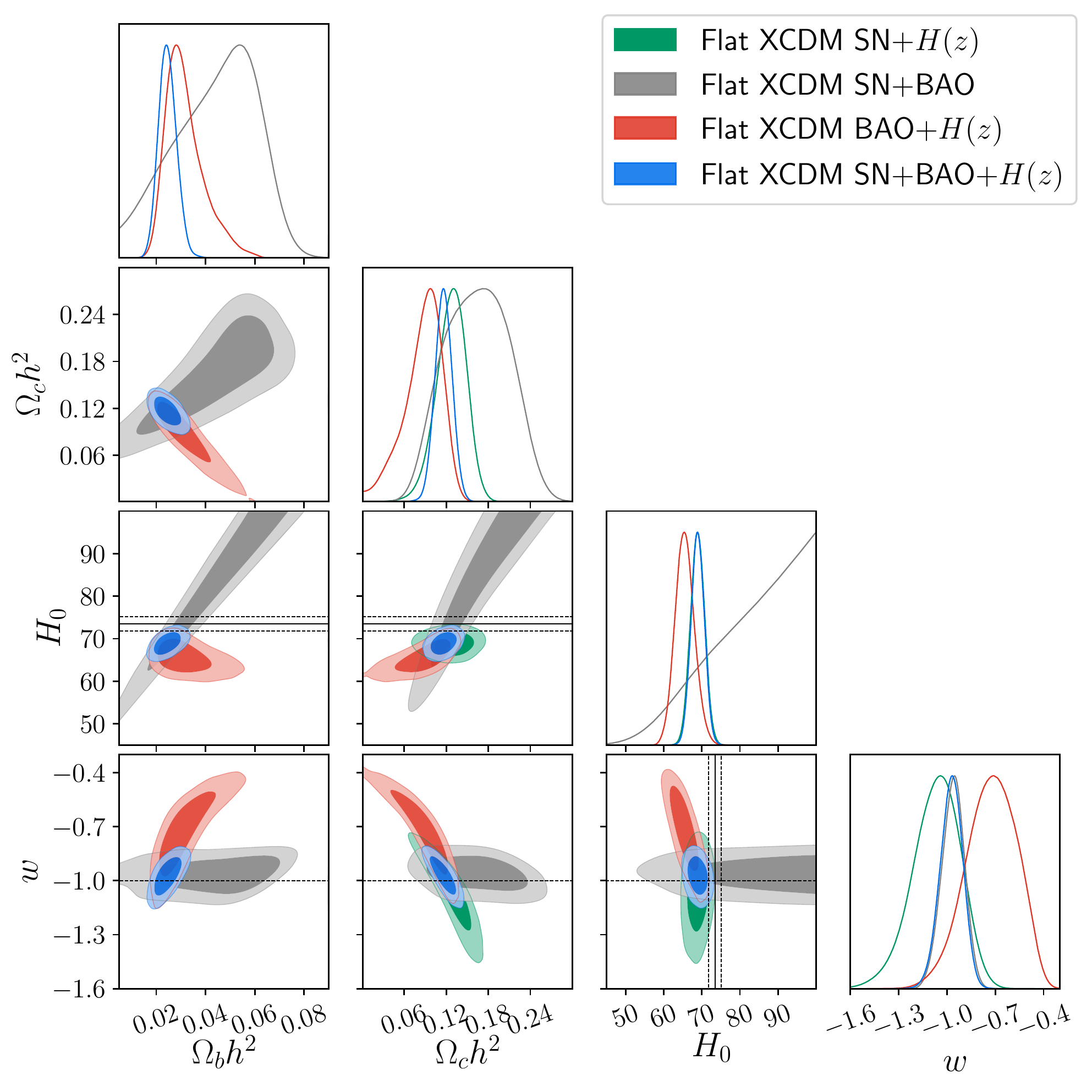}}
\mbox{\includegraphics[width=87mm]{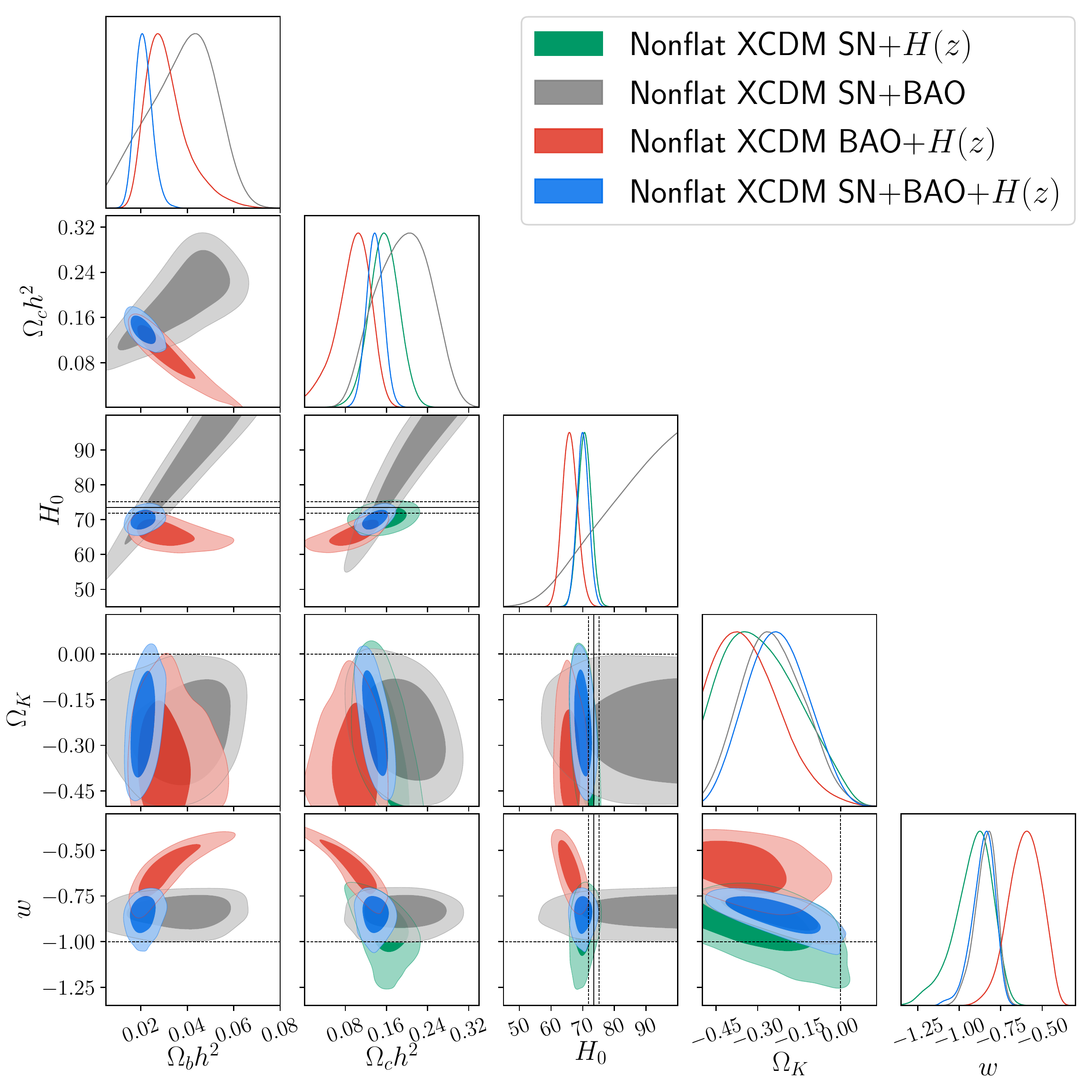}}
\caption{
Similar to Fig.\ \ref{fig:para_LCDM}, but for flat XCDM model parameters
($\Omega_b h^2$, $\Omega_c h^2$, $H_0$, $w$) in the left panel, and for nonflat
XCDM model parameters ($\Omega_b h^2$, $\Omega_c h^2$, $H_0$, $\Omega_k$, $w$) in the right panel.
The dashed lines in $w$-related plots indicate $w=-1$ (the cosmological constant). 
}
\label{fig:para_XCDM}
\end{figure*}

We modified the publicly available CAMB/COSMOMC package (version of November 
2016, \citealt{ChallinorLasenby1999,Lewisetal2000,LewisBridle2002})
to constrain the flat and nonflat $\Lambda$CDM, XCDM, and $\phi$CDM models, at
the spatially homogeneous background level, by using the SN+$H(z)$, SN+BAO, 
BAO+$H(z)$, and SN+BAO+$H(z)$ data combinations.
For the SN+$H(z)$ data combination, the model predictions are not sensitive 
to baryonic density parameter variations. In this case  $\Omega_b h^2$ is not 
constrained but instead taken to be $\Omega_b h^2 = 0.022277$, the best-fit 
value of the flat $\Lambda$CDM model constrained using Planck 2015 TT + lowP + 
lensing 
CMB data \citep{PlanckCollaboration2016}. In our analyses here we assume flat 
priors nonzero over $0.005 \le \Omega_b h^2 \le 0.1$, $0.001 \le \Omega_c h^2 \le 0.99$, $0.2 \le h \le 1.0$, $-0.5 \le \Omega_k \le 0.5$,
$-3 \le w \le 0.2$, and $0 \le \alpha \le 10$.

\section{Observational Constraints}

\begin{figure*}
\centering
\mbox{\includegraphics[width=87mm]{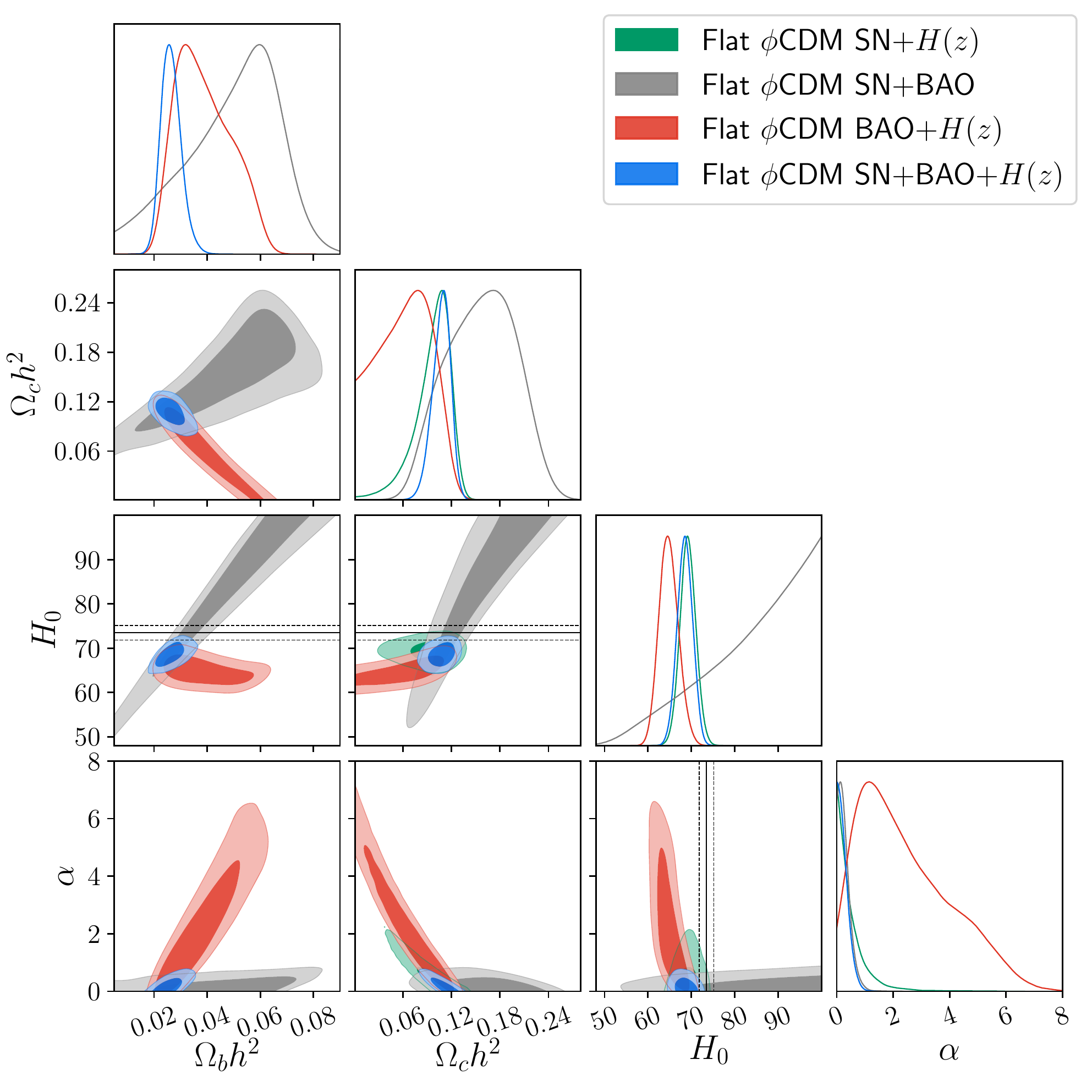}}
\mbox{\includegraphics[width=87mm]{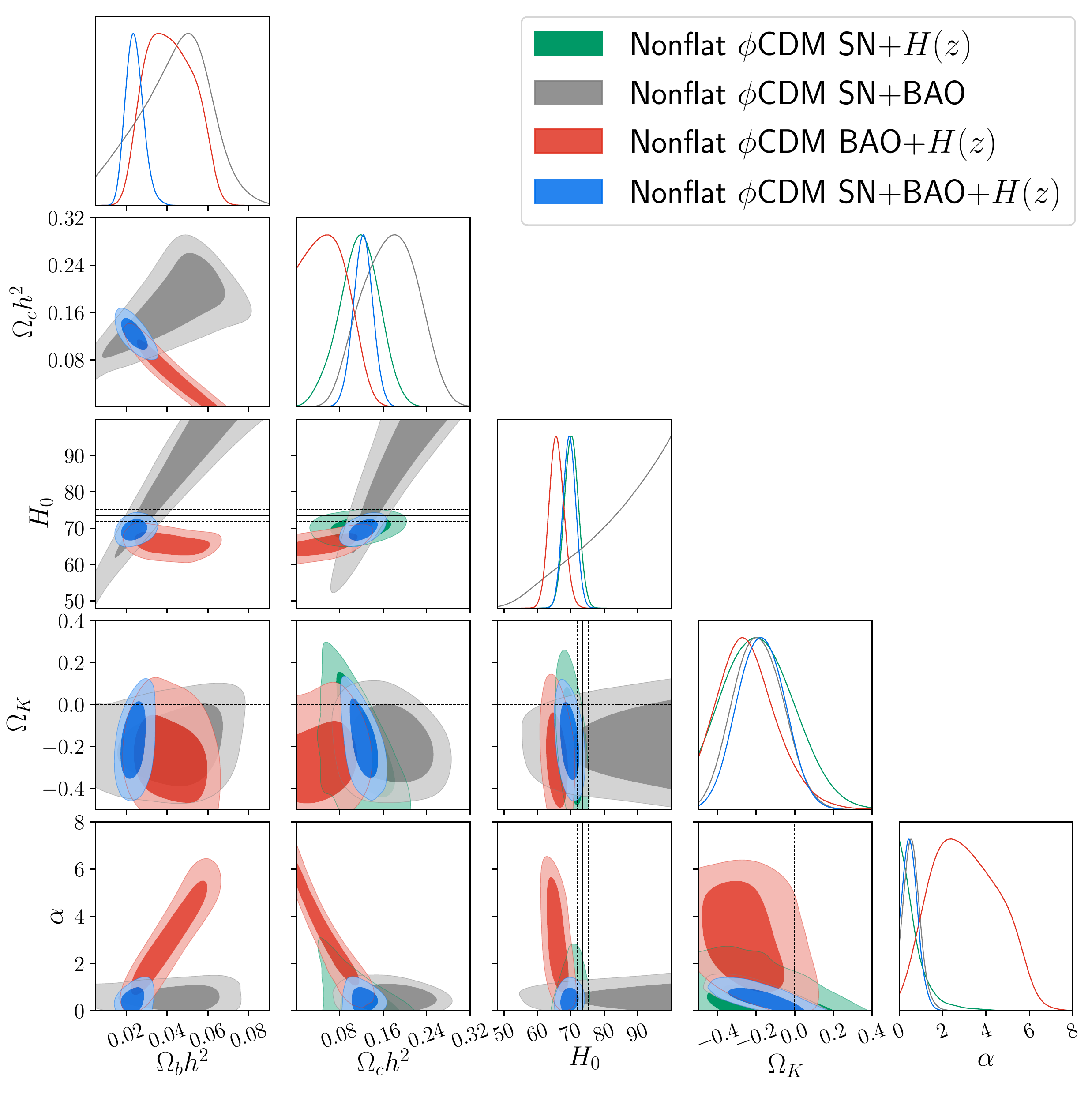}}
\caption{
Similar to Fig.\ \ref{fig:para_LCDM}, but for flat $\phi$CDM model parameters
($\Omega_b h^2$, $\Omega_c h^2$, $H_0$, $\alpha$) in the left panel, and for nonflat
$\phi$CDM model parameters ($\Omega_b h^2$, $\Omega_c h^2$, $H_0$, $\Omega_k$, $\alpha$) in the right panel.
}
\label{fig:para_phiCDM}
\end{figure*}

Figures \ref{fig:para_LCDM}--\ref{fig:para_phiCDM} show the likelihood 
distributions of model parameters of flat and nonflat $\Lambda$CDM, XCDM, 
and $\phi$CDM models, respectively. The mean and 68.3\% confidence limits 
(or 95.4\% upper limits) are summarized in Table \ref{tab:para}.

The SN+BAO data do not tightly constrain $\Omega_b h^2$, $\Omega_c h^2$, and 
especially not $H_0$, in the flat and nonflat $\Lambda$CDM, XCDM, and 
$\phi$CDM models. However, the SN+BAO data provide the most restrictive  
constraints on the dynamical dark energy parameters $w$ and $\alpha$ in the 
XCDM and $\phi$CDM models.

The results obtained using BAO+$H(z)$ data are interesting. The Hubble 
constant measured using the flat and nonflat XCDM and $\phi$CDM models are 
lower than the recent local measurement of 
$H_0 = 73.48\pm1.66$ km s$^{-1}$ Mpc$^{-1}$ \citep{Riessetal2018} by 
between 2.6$\sigma$ and 3.1$\sigma$ (of the quadrature sum of the 
two error bars), while in the flat (nonflat) $\Lambda$CDM model it is 
lower by 1.9$\sigma$ (1.3$\sigma$). In the nonflat XCDM parameterization, 
the BAO+$H(z)$ data strongly favor dark energy dynamics with $w$ deviating 
from $-1$ towards $0$ by $4.0\sigma$. For the XCDM parameterization and the 
full SN+BAO+$H(z)$ data 
set, the equation of state parameter $w$ in the flat model is measured to be 
consistent with that of the cosmological constant ($w=-1$)
while it deviates from $w=-1$ by $2.0\sigma$ in the nonflat case, which is still
significant though smaller than the $4.0\sigma$ of the BAO+$H(z)$ case.
In the nonflat $\phi$CDM model, the BAO+$H(z)$ data constraint also favors 
dark energy dynamics with $\alpha = 3.1\pm 1.5$ (a $2.1\sigma$ deviation from 
$\alpha=0$), but for the SN+BAO+$H(z)$ data combination $\alpha$ is consistent 
with zero and a cosmological constant. We note that in the nonflat $\phi$CDM 
model the CMB data alone (without lensing data) cannot tightly constrain 
$\alpha$, allowing large $\alpha \approx 10$ \citep{ParkRatra2018c},
while even the least effective combination for this here, BAO+$H(z)$, is able
to bound $\alpha < 8$, see the bottom subpanel row in the right panel of Fig.\
\ref{fig:para_phiCDM}.

For the full SN+BAO+$H(z)$ data combination, closed spatial hypersurfaces are 
favored at 1.1$\sigma$, 2.1$\sigma$, and 1.4$\sigma$ significance in the 
nonflat $\Lambda$CDM, XCDM, and $\phi$CDM models. The Planck 2015 CMB anisotropy
measurements \citep{PlanckCollaboration2016} also favor closed spatial 
hypersurfaces \citep{Oobaetal2018a, Oobaetal2018b, Oobaetal2018c}, at 
1.8$\sigma$, 1.1$\sigma$, and 1.8$\sigma$ in the $\Lambda$CDM, XCDM, 
and $\phi$CDM untilted nonflat inflation cases, 
and when combined with the SN+BAO+$H(z)$ data, as well as with growth 
factor ($f \sigma_8$) observations, they favor closed hypersurfaces at 
5.2$\sigma$, 3.4$\sigma$, and 3.1$\sigma$ significance, respectively \citep{ParkRatra2018a,ParkRatra2018b, 
ParkRatra2018c}. It is interesting, and possibly significant, that in 
Table \ref{tab:para} all three pairs of data combinations, SN+$H(z)$, SN+BAO, 
and BAO+$H(z)$, also favor closed geometries in the nonflat models, at 
between 1.0$\sigma$ and 2.9$\sigma$.\footnote{For earlier discussions of 
constraints on spatial curvature, see \citet{Farooqetal2015}, 
\citet{Chenetal2016}, \cite{YuWang2016}, \citet{LHuillierShafieloo2017}, \citet{Farooqetal2017}, \citet{WeiWu2017}, \citet{Ranaetal2017}, \citet{Yuetal2018}, 
\citet{Mitraetal2018, Mitraetal2019}, and 
\citet{Ryanetal2018, Ryanetal2019}.}

Using the SN+BAO+$H(z)$ combination, $H_0$ is measured to be  
$69.0 \pm 1.7$ ($69.8 \pm 1.8$), $68.9 \pm 1.7$ ($70.1 \pm 1.9$), and 
$68.5 \pm 1.8$ ($69.6 \pm 1.9$) km s$^{-1}$ Mpc$^{-1}$ for the flat (nonflat)
$\Lambda$CDM, XCDM, and $\phi$CDM models, respectively, These are all very 
mutually consistent and are also consistent with the most recent median 
statistics estimate of $H_0=68 \pm 2.8$ km s$^{-1}$ Mpc$^{-1}$ 
\citep{ChenRatra2011a}, which is very consistent with earlier estimates based
on median statistics \citep{Gottetal2001, Chenetal2003}.\footnote{The $H_0$ 
estimates here are consistent with many recent estimates based on non-CMB 
data \citep{LHuillierShafieloo2017,Chenetal2017,Wangetal2017,LinIshak2017,DESCollaboration2017,Yuetal2018,Haridasuetal2018a,Zhangetal2018,GomezValentAmendola2018,Haridasuetal2018b,daSilvaCavalcanti2018,Zhang2018} as well as with those
from CMB data \citep{PlanckCollaboration2018, ParkRatra2018a, ParkRatra2018b, ParkRatra2018c}.}
However, these values are a little lower than the recent local expansion 
rate measurement of $H_0 = 73.48\pm1.66$ km s$^{-1}$ Mpc$^{-1}$
\citep{Riessetal2018}\footnote{Other local expansion rate measurements 
find slightly lower $H_0$ values and slightly larger error bars
\citep{Rigaultetal2015, Zhangetal2017, Dhawanetal2017, FernandezArenasetal2018}; also see \citet{Romanetal2017}, \citet{Kimetal2018}, and 
\citet{Jonesetal2018}.} 
by between 1.9$\sigma$ and 2.0$\sigma$ for the 
flat models and between 1.3$\sigma$ and 1.5$\sigma$ for the nonflat models
(of the quadrature sum of the two error bars, in both 
cases), less discrepant than when the CMB anisotropy data is included in the 
mix \citep{ParkRatra2018a, ParkRatra2018b, ParkRatra2018c}.

It is also interesting to see the estimated values of the current matter 
density parameter (for the full non-CMB data), $\Omega_m=0.302 \pm 0.014$ ($0.321 \pm 0.022$), 
$0.297 \pm 0.019$ ($0.325 \pm 0.023$), and $0.287 \pm 0.018$ 
($0.305 \pm 0.025$) for the flat (nonflat) $\Lambda$CDM, XCDM,
and $\phi$CDM models, respectively. The flat models and both the $\phi$CDM
cases are  more consistent with the Dark Energy Survey (DES) constraint, 
$\Omega_m=0.264_{-0.019}^{+0.032}$ \citep{DESCollaboration2018} while the 
nonflat $\Lambda$CDM and XCDM model results are 1.5$\sigma$ (of the quadrature 
sum of the two error bars) larger than the DES measurement.

Table \ref{tab:chi2} summarizes the individual and total $\chi^2$ for the 
best-fit flat and nonflat $\Lambda$CDM, XCDM, and $\phi$CDM models. The 
best-fit set of parameters for each model has been determined by using 
Powell's minimization method (built into the COSMOMC program)
for finding the location of the maximum likelihood.
The $\Delta\chi^2$ of the XCDM and $\phi$CDM models denotes the 
excess $\chi^2$ relative to the $\Lambda$CDM one for the same combination of 
data sets and spatial curvature sign. The last two columns list the number 
of degrees of freedom $\nu$ 
and the reduced chi-square $\chi^2 / \nu$. The number of degrees of freedom 
is $\nu = N-n-1$, where $N$ is the number of data points and 
$n$ is the number of parameters. For example, for the nonflat $\phi$CDM model 
constrained using SN+BAO+$H(z)$ data, $N=1048+12+31=1091$ and $n=5+5=10$, 
considering the five cosmological parameters ($\Omega_b h^2$, $\Omega_c h^2$, 
$H_0$, $\alpha$, $\Omega_k$) and the five nuisance parameters of the SN sample.
Except for the case of the nonflat XCDM parameterization constrained 
using the SN+BAO+$H(z)$ data, the XCDM parametrizations fit the observations 
better than do the $\Lambda$CDM models. Furthermore, the $\phi$CDM models 
better fit the data than do the XCDM parametrizations, except for the flat 
$\phi$CDM case constrained using SN+$H(z)$ data.\footnote{The main reason 
for the smaller $\chi^2$ value in the nonflat $\phi$CDM model is that it 
fits the BAO data much better than do the $\Lambda$CDM and XCDM models.} 
However, the $\Delta\chi^2$ values are not very statistically significant.

\begin{table*}
\caption{Flat and nonflat $\Lambda\textrm{CDM}$, XCDM, and $\phi$CDM model parameter constraints from
         SN, BAO, and $H(z)$ data (mean and 68.3\% confidence limits).}
\begin{ruledtabular}
\begin{tabular}{lcccc}
  Parameter                & SN+$H(z)$            & SN+BAO               & BAO+$H(z)$           & SN+BAO+$H(z)$        \\[+0mm]
 \hline \\[-2mm]
 \multicolumn{5}{c}{Flat $\Lambda\textrm{CDM}$ model} \\
  \hline \\[-2mm]
  $\Omega_b h^2$           & $\ldots$             & $0.043 \pm 0.014$    & $0.02310 \pm 0.0030$ & $0.0235 \pm 0.0028$  \\[+1mm]
  $\Omega_c h^2$           & $0.1205 \pm 0.0088$  & $0.174 \pm 0.042$    & $0.1212  \pm 0.0086$ & $0.1197 \pm 0.0074$  \\[+1mm]
  $H_0$ [km s$^{-1}$ Mpc$^{-1}$] & $69.1 \pm 1.8$ & $>61.4$ (95.4\% C.L.)& $68.8    \pm 1.8$    & $69.0   \pm 1.7$     \\[+1mm]
  $\Omega_m$               & $0.301 \pm 0.020$    & $0.302 \pm 0.015$    & $0.306   \pm 0.019$  & $0.302  \pm 0.014$   \\[+1mm]
 \hline \\[-2mm]
 \multicolumn{5}{c}{Nonflat $\Lambda\textrm{CDM}$ model} \\
  \hline \\[-2mm]
  $\Omega_b h^2$           & $\ldots$             & $0.036 \pm 0.015$    & $0.0205 \pm 0.0037$  & $0.0207 \pm 0.0036$  \\[+1mm]
  $\Omega_c h^2$           & $0.168 \pm 0.027$    & $0.187 \pm 0.054$    & $0.137  \pm 0.017$   & $0.135  \pm 0.016$   \\[+1mm]
  $H_0$ [km s$^{-1}$ Mpc$^{-1}$] & $70.2 \pm 2.0$ & $>57.0$ (95.4\% C.L.)& $70.1   \pm 2.1$     & $69.8   \pm 1.8$     \\[+1mm]
  $\Omega_k$               & $-0.23 \pm 0.12$     & $-0.066 \pm 0.066$   & $-0.086 \pm 0.078$   & $-0.072 \pm 0.065$   \\[+1mm]
  $\Omega_m$               & $0.387 \pm 0.049$    & $0.319 \pm 0.023$    & $0.322  \pm 0.023$   & $0.321  \pm 0.022$   \\[+1mm]
 \hline 
 \hline \\[-2mm]
 \multicolumn{5}{c}{Flat XCDM parameterization} \\
  \hline \\[-2mm]
  $\Omega_b h^2$           & $\ldots$             & $0.044 \pm 0.016$    & $0.0317 \pm 0.0080$  & $0.0246 \pm 0.0036$  \\[+1mm]
  $\Omega_c h^2$           & $0.127 \pm 0.022$    & $0.164 \pm 0.045$    & $0.087  \pm 0.027$   & $0.116  \pm 0.012$  \\[+1mm]
  $H_0$ [km s$^{-1}$ Mpc$^{-1}$] & $68.9 \pm 1.8$ & $>57.0$ (95.4\% C.L.)& $65.5   \pm 2.5$     & $68.9   \pm 1.7$     \\[+1mm]
  $w$                      & $-1.07 \pm 0.15$     & $-0.963 \pm 0.070$   & $-0.72  \pm 0.16$    & $-0.973 \pm 0.071$   \\[+1mm]
  $\Omega_m$               & $0.316 \pm 0.048$    & $0.295 \pm 0.019$    & $0.276  \pm 0.035$   & $0.297  \pm 0.019$   \\[+1mm]
 \hline \\[-2mm]
 \multicolumn{5}{c}{Nonflat XCDM parameterization} \\
  \hline \\[-2mm]
  $\Omega_b h^2$           & $\ldots$             & $0.038 \pm 0.014$    & $0.0313 \pm 0.0093$  & $0.0212 \pm 0.0037$  \\[+1mm]
  $\Omega_c h^2$           & $0.155 \pm 0.029$    & $0.193 \pm 0.051$    & $0.095  \pm 0.033$   & $0.138  \pm 0.016$  \\[+1mm]
  $H_0$ [km s$^{-1}$ Mpc$^{-1}$] & $70.6 \pm 2.1$ & $>59.6$ (95.4\% C.L.)& $66.0   \pm 2.4$     & $70.1   \pm 1.9$     \\[+1mm]
  $\Omega_k$               & $-0.28 \pm 0.13$     & $-0.24 \pm 0.11$     & $-0.32  \pm 0.11$    & $-0.23  \pm 0.11$   \\[+1mm]
  $w$                      & $-0.92 \pm 0.12$     & $-0.841 \pm 0.066$   & $-0.604 \pm 0.099$   & $-0.856 \pm 0.071$   \\[+1mm]
  $\Omega_m$               & $0.358 \pm 0.056$    & $0.322 \pm 0.022$    & $0.291  \pm 0.044$   & $0.325  \pm 0.023$   \\[+1mm]
 \hline 
 \hline \\[-2mm]
 \multicolumn{5}{c}{Flat $\phi$CDM model} \\
  \hline \\[-2mm]
  $\Omega_b h^2$           & $\ldots$             & $0.049 \pm 0.017$    & $0.039 \pm 0.010$    & $0.0264 \pm 0.0038$  \\[+1mm]
  $\Omega_c h^2$           & $0.097 \pm 0.023$    & $0.157 \pm 0.042$    & $0.062  \pm 0.032$   & $0.108  \pm 0.011$  \\[+1mm]
  $H_0$ [km s$^{-1}$ Mpc$^{-1}$] & $69.2 \pm 1.8$ & $>57.2$ (95.4\% C.L.)& $64.8   \pm 2.2$     & $68.5   \pm 1.8$     \\[+1mm]
  $\alpha$ [95.4\% C.L.]   & $ < 2.2$             & $ < 0.82$            & $ < 6.0$ [$2.5 \pm 1.6$ (68.3\% C.L.)] & $ < 0.73$            \\[+1mm]
  $\Omega_m$               & $0.250 \pm 0.049$    & $0.284 \pm 0.019$    & $0.241  \pm 0.045$   & $0.287  \pm 0.018$   \\[+1mm]
 \hline \\[-2mm]
 \multicolumn{5}{c}{Nonflat $\phi$CDM model} \\
  \hline \\[-2mm]
  $\Omega_b h^2$           & $\ldots$             & $0.044 \pm 0.017$    & $0.041 \pm 0.011$    & $0.0240 \pm 0.0043$  \\[+1mm]
  $\Omega_c h^2$           & $0.116 \pm 0.037$    & $0.173 \pm 0.049$    & $0.060 \pm 0.035$    & $0.123  \pm 0.018$  \\[+1mm]
  $H_0$ [km s$^{-1}$ Mpc$^{-1}$] & $70.2 \pm 2.1$ & $>57.4$ (95.4\% C.L.)& $65.8   \pm 2.2$     & $69.6   \pm 1.9$     \\[+1mm]
  $\Omega_k$               & $-0.17 \pm 0.17$     & $-0.19 \pm 0.13$     & $-0.24  \pm 0.15$    & $-0.17  \pm 0.12$   \\[+1mm]
  $\alpha$ [95.4\% C.L.]   & $ < 2.8$             & $ < 1.5$             & $3.1 \pm 1.5$ (68.3\% C.L.)  & $ < 1.3$            \\[+1mm]
  $\Omega_m$               & $0.283 \pm 0.072$    & $0.300 \pm 0.025$    & $0.235 \pm 0.048$    & $0.305  \pm 0.025$   \\[+0mm]
\end{tabular}
\\[+1mm]
Note: $\Omega_m$ is a derived parameter.
\end{ruledtabular}
\label{tab:para}
\end{table*}

\begin{table*}
\caption{Individual and total $\chi^2$ values for the best-fit flat and nonflat $\Lambda$CDM, XCDM, and $\phi$CDM models.}
\begin{ruledtabular}
\begin{tabular}{lcccrrrr}
    Data sets   &   $\chi_{\textrm{SN}}^2$  & $\chi_{\textrm{BAO}}^2$  &  $\chi_{H(z)}^2$   &  Total $\chi^2$      & $\Delta\chi^2$  & $\nu$   &  $\chi^2 / \nu $   \\[+0mm]
\hline \\[-2mm]
\multicolumn{8}{c}{Flat $\Lambda$CDM model} \\
\hline \\[-2mm]
   SN+$H(z)$      &  1035.98 &          &    14.61 &  1050.59  &           &  1070   &  0.9819 \\[+1mm]
   SN+BAO         &  1035.99 &    10.06 &          &  1046.05  &           &  1051   &  0.9953 \\[+1mm]
   BAO+$H(z)$     &          &    10.03 &    14.58 &    24.61  &           &    39   &  0.6310 \\[+1mm]
   SN+BAO+$H(z)$  &  1036.00 &    10.03 &    14.61 &  1060.64  &           &  1082   &  0.9803 \\[+0mm]
\hline\\[-2mm]
\multicolumn{8}{c}{Nonflat $\Lambda$CDM model} \\
\hline \\[-2mm]
   SN+$H(z)$      &  1035.88 &          &    14.56 &  1050.44  &           &  1069   &  0.9826 \\[+1mm]
   SN+BAO         &  1036.10 &    10.04 &          &  1046.13  &           &  1050   &  0.9963 \\[+1mm]
   BAO+$H(z)$     &          &    10.29 &    14.97 &    25.26  &           &    38   &  0.6647 \\[+1mm]
   SN+BAO+$H(z)$  &  1036.06 &    10.02 &    14.58 &  1060.66  &           &  1081   &  0.9812 \\[+0mm]
 \hline 
\hline\\[-2mm]
\multicolumn{8}{c}{Flat XCDM parameterization} \\
\hline \\[-2mm]
   SN+$H(z)$      &  1035.93 &          &    14.43 &  1050.37  & $-0.22$   &  1069   &  0.9826 \\[+1mm]
   SN+BAO         &  1036.12 &     9.73 &          &  1045.84  & $-0.21$   &  1050   &  0.9960 \\[+1mm]
   BAO+$H(z)$     &          &     7.03 &    14.98 &    22.01  & $-2.60$   &    38   &  0.5792 \\[+1mm]
   SN+BAO+$H(z)$  &  1036.25 &     9.57 &    14.67 &  1060.49  & $-0.15$   &  1081   &  0.9810 \\[+0mm]
\hline\\[-2mm]
\multicolumn{8}{c}{Nonflat XCDM parameterization} \\
\hline \\[-2mm]
   SN+$H(z)$      &  1035.90 &          &    14.51 &  1050.41  & $-0.03$   &  1068   &  0.9835 \\[+1mm]
   SN+BAO         &  1036.03 &     9.98 &          &  1046.00  & $-0.13$   &  1049   &  0.9971 \\[+1mm]
   BAO+$H(z)$     &          &    10.30 &    14.71 &    25.02  & $-0.24$   &    37   &  0.6762 \\[+1mm]
   SN+BAO+$H(z)$  &  1036.15 &     9.81 &    14.81 &  1060.76  & $+0.10$   &  1080   &  0.9822 \\[+0mm]
 \hline 
\hline\\[-2mm]
\multicolumn{8}{c}{Flat $\phi$CDM model} \\
\hline \\[-2mm]
   SN+$H(z)$      &  1035.98 &          &    14.61 &  1050.59  &  $0.00$   &  1069   &  0.9828 \\[+1mm]
   SN+BAO         &  1036.38 &     9.37 &          &  1045.74  & $-0.31$   &  1050   &  0.9959 \\[+1mm]
   BAO+$H(z)$     &          &     7.00 &    14.89 &    21.88  & $-2.73$   &    38   &  0.5758 \\[+1mm]
   SN+BAO+$H(z)$  &  1036.36 &     9.39 &    14.68 &  1060.43  & $-0.21$   &  1081   &  0.9810 \\[+0mm]
\hline\\[-2mm]
\multicolumn{8}{c}{Nonflat $\phi$CDM model} \\
\hline \\[-2mm]
   SN+$H(z)$      &  1035.84 &          &    14.54 &  1050.38  & $-0.06$   &  1068   &  0.9835 \\[+1mm]
   SN+BAO         &  1036.07 &     8.02 &          &  1044.09  & $-2.04$   &  1049   &  0.9953 \\[+1mm]
   BAO+$H(z)$     &          &     4.14 &    15.05 &    19.19  & $-6.07$   &    37   &  0.5186 \\[+1mm]
   SN+BAO+$H(z)$  &  1036.26 &     7.86 &    15.14 &  1059.27  & $-1.39$   &  1080   &  0.9808 \\[+0mm]
\end{tabular}
\\[+1mm]
Note: $\Delta\chi^2$ of the XCDM and $\phi$CDM models represent the excess value relative to $\chi^2$
of the corresponding $\Lambda\textrm{CDM}$ model for the same combination of data sets and spatial curvature sign.
\end{ruledtabular}
\label{tab:chi2}
\end{table*}

\begin{figure*}
\centering
\mbox{\includegraphics[width=40mm]{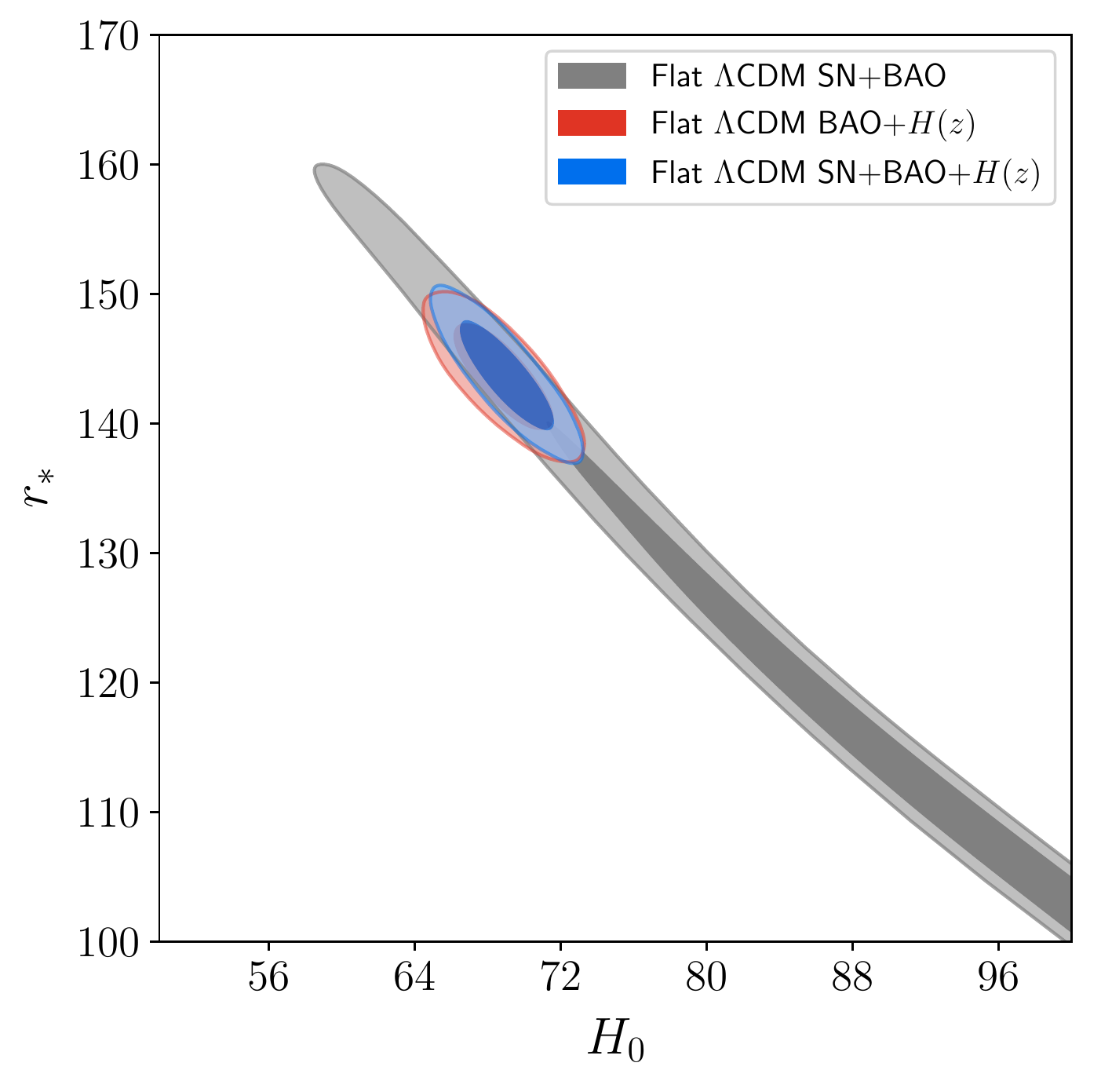}}
\mbox{\includegraphics[width=40mm]{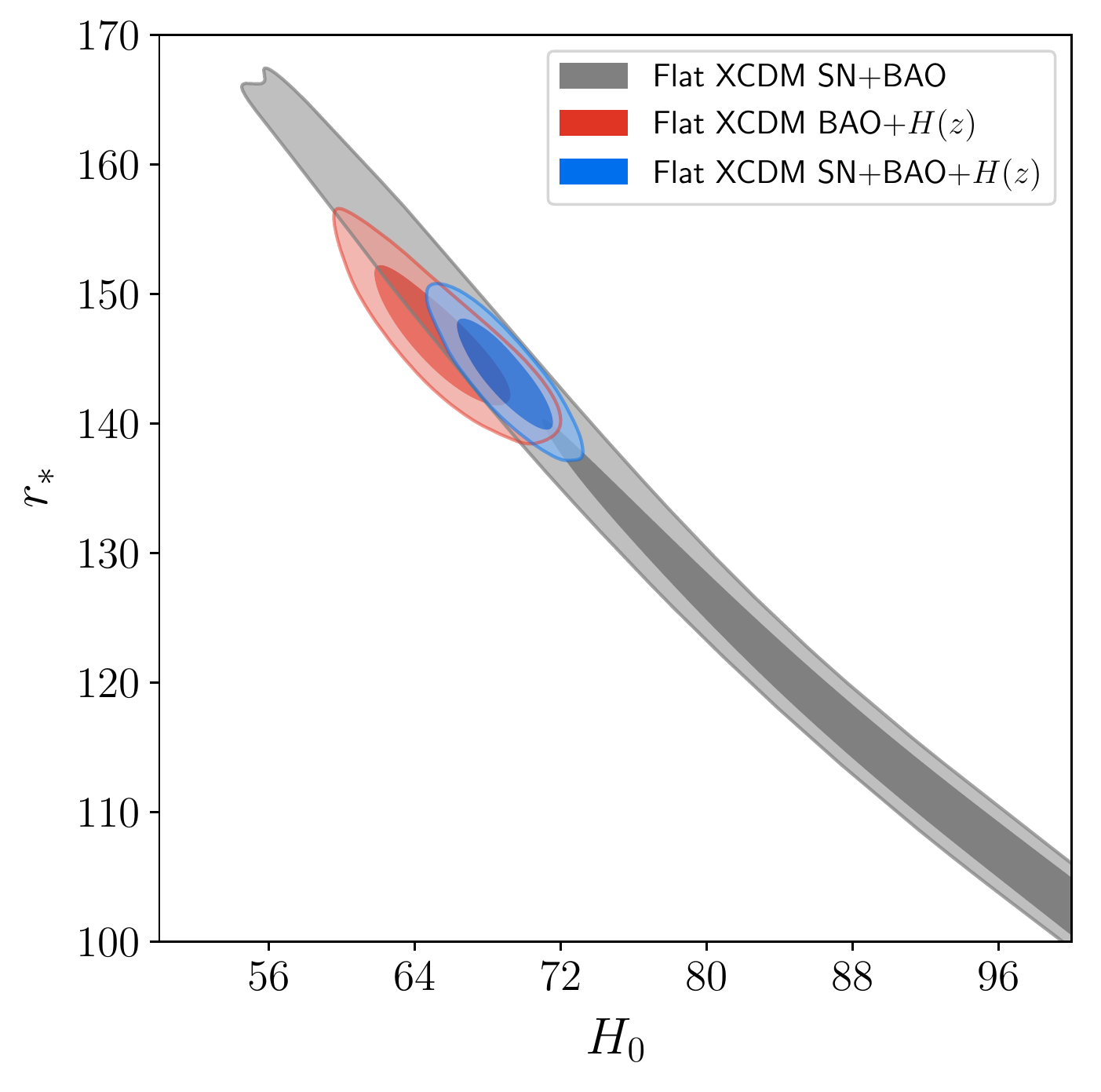}}
\mbox{\includegraphics[width=40mm]{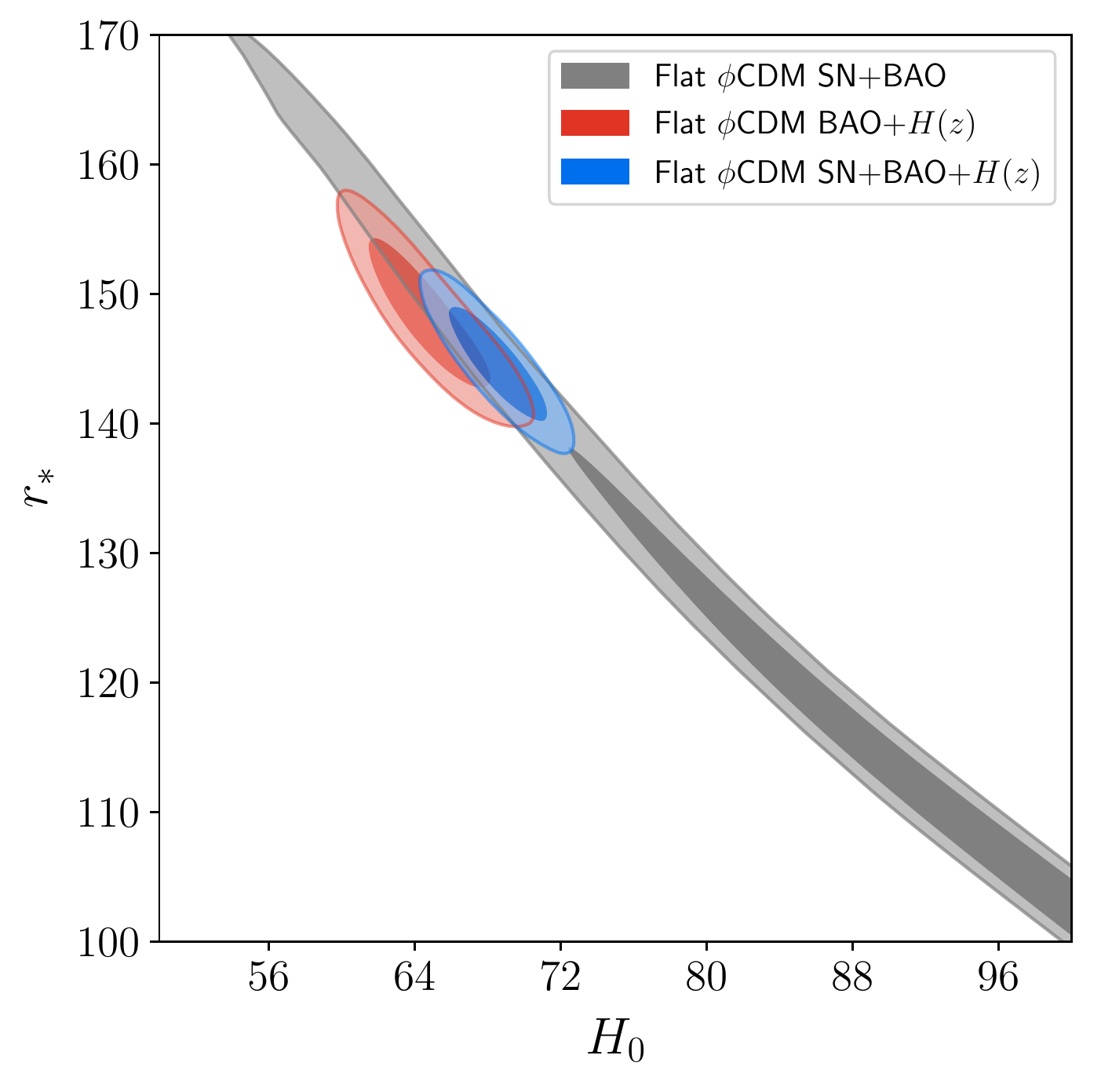}} \\
\mbox{\includegraphics[width=40mm]{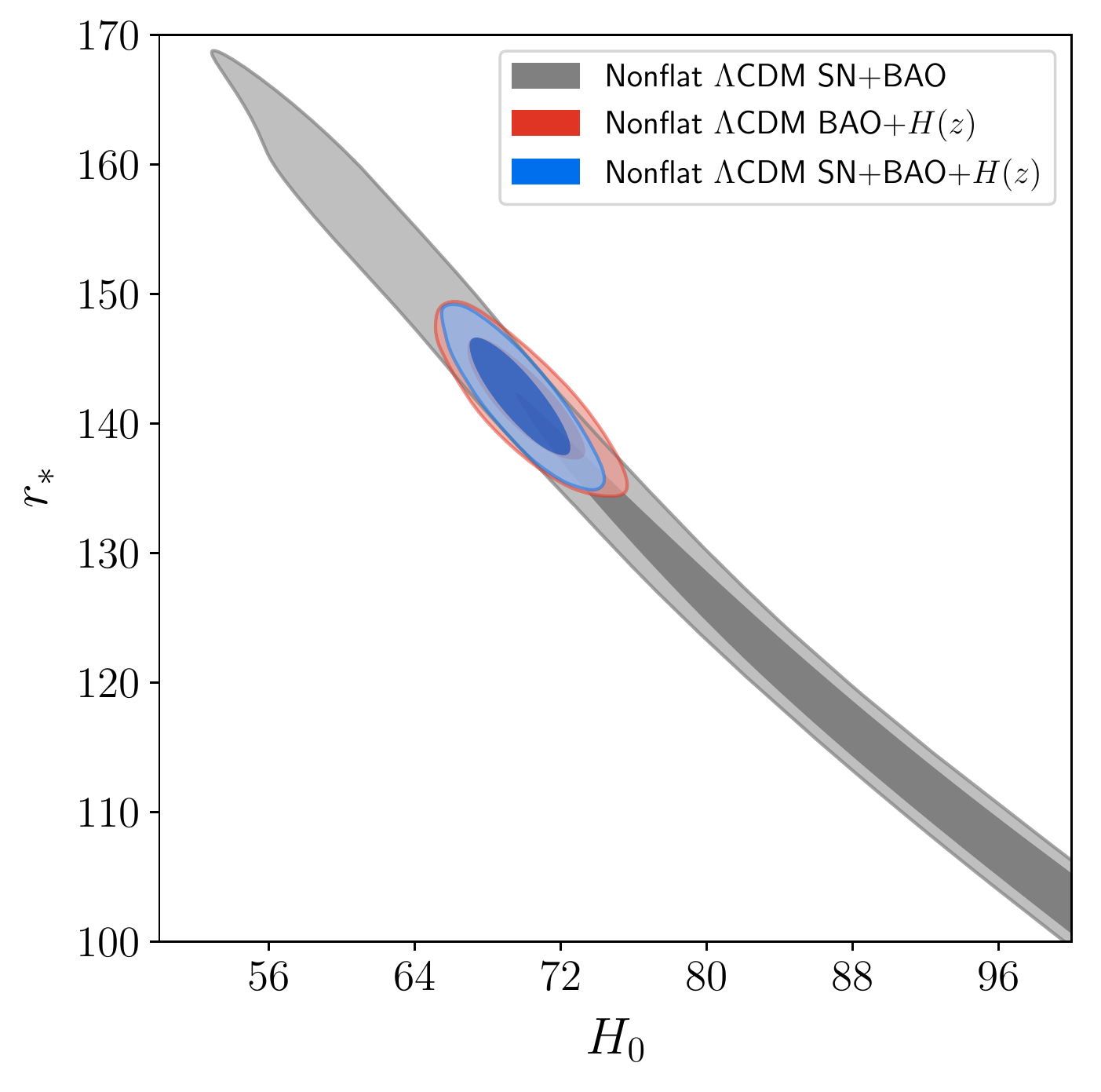}}
\mbox{\includegraphics[width=40mm]{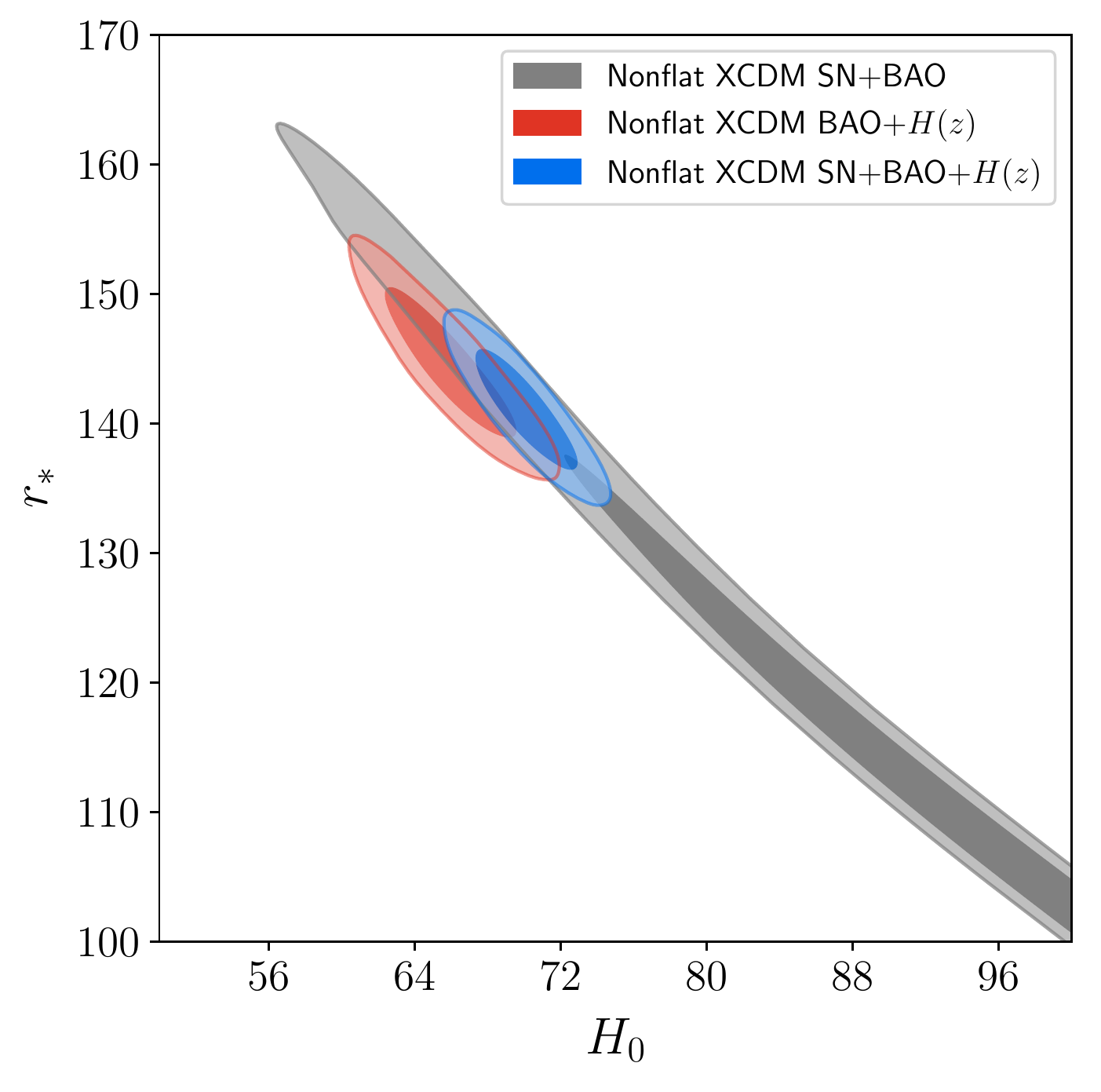}}
\mbox{\includegraphics[width=40mm]{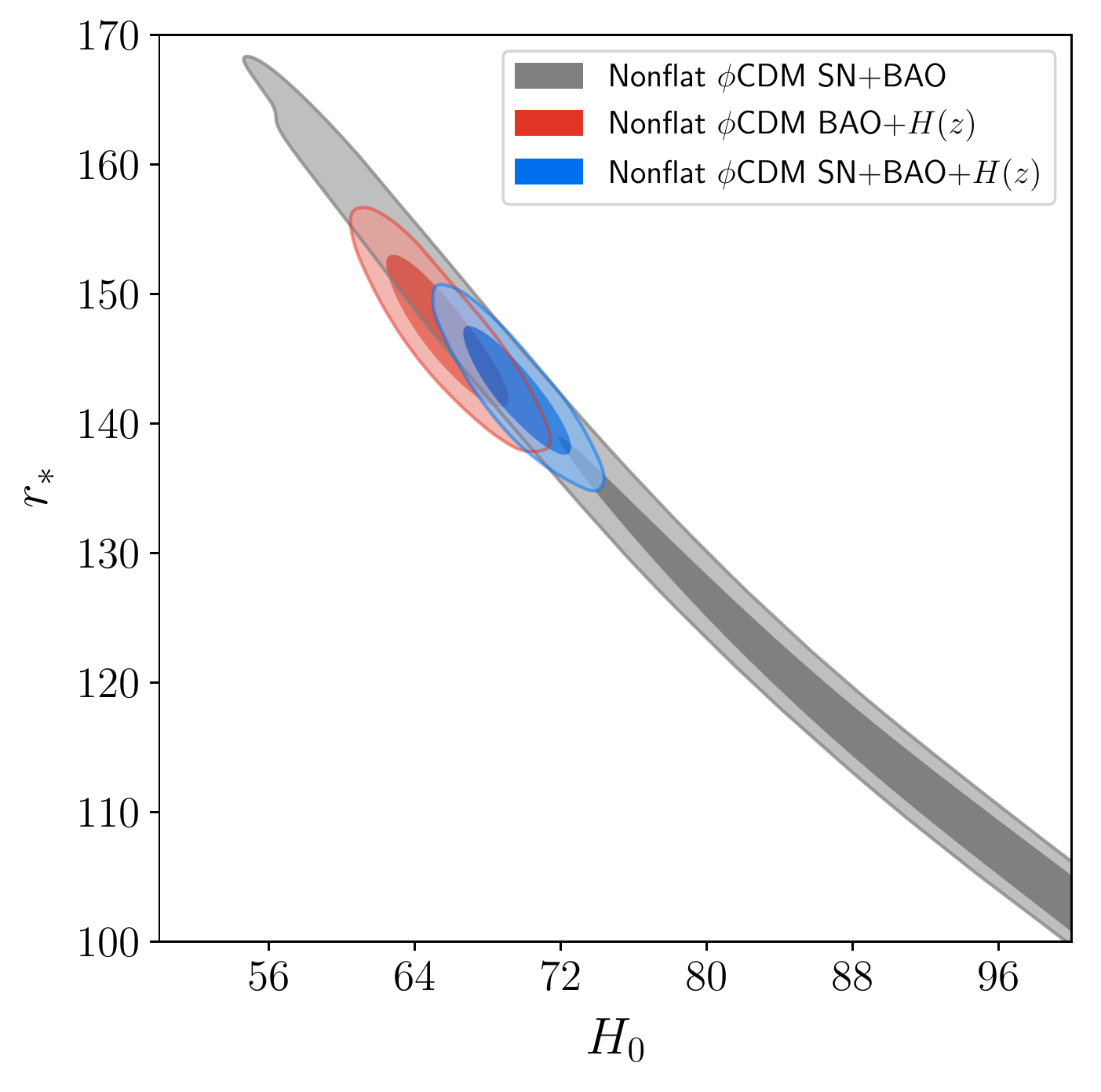}}
\caption{
Hubble constant ($H_0$) versus sound horizon size at recombination ($r_*$) in the flat (top) and nonflat (bottom panels) $\Lambda$CDM, XCDM, $\phi$CDM models constrained by SN+BAO, BAO+$H(z)$, and SN+BAO+$H(z)$ data sets.
}
\label{fig:H0rs_LCDM}
\end{figure*}

Comparing the results for SN+$H(z)$ and SN+BAO data in Figs. \ref{fig:para_LCDM}--\ref{fig:para_phiCDM},
we see that the BAO data are less restrictive than the Hubble parameter data in the parameter estimation.
Especially, SN+BAO data do not provide a tight constraint on $H_0$, allowing extreme values of
the Hubble constant $H_0 > 90~\textrm{km}\ \textrm{s}^{-1} \textrm{Mp}c^{-1}$. 
This seems to be in contradiction to the recent estimation of Hubble constant using the inverse distance ladder method
\citep{Macaulayetal2019,Aubourg2015}, where the Hubble constant has been tightly constrained by the SN and BAO data with a reasonable prior
on the sound horizon size at recombination ($r_*$) based on the CMB data. 
In our analysis, however, we do not assume any prior on the sound horizon size because we aim to see
how the cosmological parameters of the dark energy models are constrained without relying on the CMB data.
Figure \ref{fig:H0rs_LCDM} shows the relation between the Hubble constant ($H_0$)
and the sound horizon size at recombination ($r_*$) in the six models considered here. As expected, the case of SN+BAO data shows
strong correlation between $H_0$ and $r_*$. For a higher value of Hubble constant, the lower sound horizon size is favored.
We note that such a low value of sound horizon (e.g., $r_*\simeq 100$ Mpc) is certainly unrealistic in most cosmological models.
However, we emphasize that adding Hubble parameter measurements to our analysis provide a very tight constraint on the sound horizon size.
For SN+BAO+$H(z)$ data set, $r_*=143.7 \pm 2.8$ Mpc ($142.0 \pm 3.0$ Mpc) in flat (nonflat) $\Lambda$CDM model, 
$r_*=143.8 \pm 2.8$ Mpc ($141.1 \pm 3.1$ Mpc) in flat (nonflat) XCDM model, and $r_*=144.6 \pm 2.9$ Mpc ($142.5 \pm 3.2$ Mpc) in 
flat (nonflat) $\phi$CDM model. These values are consistent with $r_*=144.43 \pm 0.26$ Mpc of the flat $\Lambda$CDM model constrained with
the Planck 2018 data (TT,TE,EE+lowE+lensing, \citealt{PlanckCollaboration2018}).

\section{Summary}

We have used Type Ia supernova apparent magnitude, baryon acoustic 
oscillation distance, and Hubble parameter measurements to constrain parameters 
of the flat and nonflat $\Lambda$CDM, XCDM, and $\phi$CDM models.

Our main results, in summary, are:
\begin{itemize}
\item These data favor closed spatial hypersurfaces at 1.1$\sigma$ to 
2.1$\sigma$, depending on the nonflat model.
\item These data do not rule out dark energy dynamics.
\item These data favor a smaller Hubble constant than the recent local 
expansion rate measurement of $H_0 = 73.48\pm1.66$ km s$^{-1}$ Mpc$^{-1}$ 
\citep{Riessetal2018}  at 1.3$\sigma$ to 2.0$\sigma$, depending on model.
\end{itemize}

These results are consistent with those that follow from similar analyses of 
CMB anisotropy data in untilted nonflat inflation models, and consequently 
joint analyses of CMB and non-CMB data reinforce the above findings 
\citep{ParkRatra2018a,ParkRatra2018b,ParkRatra2018c}. 

%
%
\acknowledgements{
C.-G.P.\ was supported by the Basic Science Research Program through the National Research Foundation of Korea (NRF)
funded by the Ministry of Education (No.\ 2017R1D1A1B03028384). B.R.\ was supported in part by DOE
grant DE-SC0019038.
}

%
%

\def\and{{and }}
\bibliographystyle{yahapj}


\end{document}